\documentclass[pdflatex,sn-mathphys-ay]{sn-jnl} 

\usepackage{graphicx}
\usepackage{multirow}
\usepackage{amsmath,amssymb,amsfonts}
\usepackage{amsthm}
\usepackage{mathrsfs}
\usepackage[title]{appendix}
\usepackage{xcolor}
\usepackage{textcomp}
\usepackage{manyfoot}
\usepackage{booktabs}
\usepackage{algorithm}
\usepackage{algorithmicx}
\usepackage{algpseudocode}
\usepackage{listings}
\usepackage{lscape}
\usepackage{natbib}

\theoremstyle{thmstyleone}%
\theoremstyle{thmstyletwo}%

\theoremstyle{thmstylethree}%

\begin{document}

\title[Article Title]{Deep Knowledge Tracing for Personalized Adaptive
Learning at Historically Black Colleges and
Universities}


\author{\fnm{Ming-Mu} \sur{Kuo}}\email{mmkuo@pvamu.edu}
\equalcont{These authors contributed equally to this work.}

\author{\fnm{Xiangfang} \sur{Li}}\email{xili@pvamu.edu}
\equalcont{These authors contributed equally to this work.}

\author{\fnm{Lijun} \sur{Qian}}\email{liqian@pvamu.edu}
\equalcont{These authors contributed equally to this work.}

\author{\fnm{Pamela} \sur{Obiomon}}\email{phobiomon@pvamu.edu}
\equalcont{These authors contributed equally to this work.}

\author*{\fnm{Xishuang} \sur{Dong}}\email{xidong@pvamu.edu}
\equalcont{These authors contributed equally to this work.}

\affil{\orgdiv{CREDIT Center and ECE Department}, \orgname{Prairie View A\&M University}, \orgaddress{\street{700 University Drive}, \city{Prairie View}, \postcode{77446}, \state{TX}, \country{USA}}}


\abstract{Personalized adaptive learning (PAL) stands out by closely monitoring individual students' progress and tailoring their learning paths to their unique knowledge and needs. A crucial technique for effective PAL implementation is knowledge tracing, which models students' evolving knowledge to predict their future performance. Recent advancements in deep learning have significantly enhanced knowledge tracing through Deep Knowledge Tracing (DKT). However, there is limited research on DKT for Science, Technology, Engineering, and Math (STEM) education at Historically Black Colleges and Universities (HBCUs). This study builds a comprehensive dataset to investigate DKT for implementing PAL in STEM education at HBCUs, utilizing multiple state-of-the-art (SOTA) DKT models to examine knowledge tracing performance. The dataset includes $352,148$ learning records for $17,181$ undergraduate students across eight colleges at Prairie View A\&M University (PVAMU). The SOTA DKT models employed include DKT, DKT+, DKVMN, SAKT, and KQN. Experimental results demonstrate the effectiveness of DKT models in accurately predicting students’ academic outcomes. Specifically, the SAKT and KQN models outperform others in terms of accuracy and AUC. These findings have significant implications for faculty members and academic advisors, providing valuable insights for identifying students at risk of academic underperformance before the end of the semester. Furthermore, this allows for proactive interventions to support students' academic progress, potentially enhancing student retention and graduation rates.}

\keywords{Historically Black Colleges and Universities, STEM Education, Personalized Adaptive Learning, Knowledge Tracing}



\maketitle

\section*{Introduction}
In contemporary educational settings, personalized adaptive learning (PAL) has emerged as a dynamic approach to cater to the diverse needs and learning styles of students. This pedagogical framework tailors instructional content, pace, and learning experiences to the individual learner, leveraging technology and data analytics to facilitate targeted interventions \citep{bajaj2018smart}. In particular, PAL systems utilize artificial intelligence techniques to analyze student performance data, enabling educators to gain insights into each learner's strengths, weaknesses, and areas for improvement \citep{zine2019comparative,yagci2022educational,waheed2020predicting}. By providing customized learning pathways and real-time feedback, these systems aim to optimize student engagement, comprehension, and overall academic achievement \citep{essa2023personalized}.
Deep Knowledge Tracing (DKT) emerges as a crucial component for implementing PAL systems. It represents a paradigm shift from traditional assessment methods, offering a sophisticated approach to modeling and predicting students' mastery of specific concepts or skills over time \citep{piech2015deep}. Unlike conventional methods that rely solely on grades or assessments, DKT leverages deep learning algorithms to analyze sequential student interactions with learning materials, including exercises, quizzes, and assignments. By capturing the dynamic and temporal aspects of learning, DKT predicts future learning outcomes, aiding educators in understanding and addressing students' individual progress and needs. This provides a more intricate comprehension of students' knowledge acquisition processes, facilitating timely interventions and personalized support \citep{liu2023survey}. This nuanced understanding enables tailored support, fostering a more effective and responsive learning environment.

Despite the advancements in DKT, there remains a gap in its investigation and application within specific educational contexts, particularly at Historically Black Colleges and Universities (HBCUs) \citep{Pokrajac2016}. These institutions play a vital role in fostering academic excellence and opportunity for underrepresented minority students, yet there is a lack of comprehensive research on the effectiveness and implementation of DKT within HBCU settings. Addressing this gap is crucial for advancing the understanding of knowledge tracing in diverse educational environments and ensuring equitable access to effective learning technologies.

This study builds a comprehensive dataset to investigate DKT for implementing PAL in STEM education at HBCUs. It utilizes multiple state-of-the-art (SOTA) DKT models, including DKT, DKT+, DKVMN, SAKT, and KQN, to examine knowledge tracing performance. The dataset comprises $352,148$ learning records for $17,181$ undergraduate students across eight colleges at Prairie View A\&M University (PVAMU), with a specialized focus on College of Engineering majors. The study centers on the unique challenges and opportunities within engineering education, detailing the tailored data collection processes, model development strategies, and evaluation procedures employed. Experimental results demonstrate the effectiveness of these DKT models in accurately predicting students' academic outcomes. Specifically, the SAKT and KQN models outperform others in terms of accuracy and AUC. Additionally, the implications of these findings for personalized adaptive learning initiatives and educational practices in HBCU settings are discussed.

The contributions of this study are summarized as follows:

\begin{itemize}
\item To the best of our knowledge, this is the first dataset to investigate DKT at HBCUs, exploring the potentials of PAL. It enables the capture, analysis, and interpretation of data relevant to STEM education at HBCUs using advanced knowledge tracing techniques to implement PAL. 

\item This study evaluates the effectiveness and limitations of current SOTA DKT models in predicting student performance and identifying at-risk individuals early in the semester for HBCUs. Moreover, it addresses the gap in the investigation and application of DKT within the specific educational contexts of HBCUs.
 It not only broadens the scope of DKT applications but also provides valuable insights into effective implementation strategies and implications for future research and practice in diverse educational settings.
 
\end{itemize}

\section*{Data collection and preparation}

\subsection*{Data Introduction}
Educational institutions routinely maintain comprehensive electronic records of student information, encompassing diverse data types and volumes, ranging from demographic details to academic achievements. This study collected a real dataset of undergraduate students from the Student Information System (SIS) at PVAMU, one HBCU in Texas. The data spans four years, from fall 2020 to summer 2023. The collected data includes five essential elements: Academic Year, Universal ID, Course Subject, Course Level, and Pass/Fail grades. These features are strictly academic and relate to students' grades in their courses over the academic years. Table \ref{tab_cp2} provides a few examples from the dataset.

\begin{table}[!ht]
	\caption{ Data samples }
\centering 
\begin{tabular}{ccccc}
\hline
\textbf{Academic Year}  & \textbf{Universal ID}   & \textbf{Course Subject} & \textbf{Course Level}  & \textbf{Pass/Fail} \\ \hline

2020     & 5517806   & BIOL    & 1000   & 1                             \\ \hline
2020     & 5578432   & ENGL    & 1000   & 1                             \\ \hline
2021     & 5626380   & MATH    & 2000   & 0                             \\ \hline
2022     & 5966264   & ELEG    & 2000   & 1                             \\ \hline
2023     & 4929554   & HLTH    & 4000   & 0                             \\ \hline
2023     & 4929554   & KINE    & 4000   & 1                             \\ \hline
...     & ...   & ...    & ...   & ...                           \\ \hline
\end{tabular}
	\label{tab_cp2}
\end{table}

where ``Academic Year" denotes the specific academic year during which the course was undertaken, providing temporal context to the data. The original student ID was substituted with a randomly generated universal ID to ensure the anonymity of individual students. The course subject consists of the four characters associated with the course. The course level is determined by the thousands digit in the course number. Freshman courses commence at the one thousand level, sophomore courses at the two thousand level, junior courses at the three thousand level, senior courses at the four thousand level, graduate-level courses at five thousand and above, and development courses initiate at the zero level. The Pass/Fail data is binary, denoting a pass with the value 1 or a failure with the value 0. This study considers grades of A, B, C, and 'Credit/Pass' as indicative of passing the course, while grades of D, F, withdraw and 'No credit/did not pass' signify failure.

\subsection*{Data Preparation}
To prepare data for DKT, a few steps of data preprocessing are completed: 1) Data records lacking grades, incomplete courses, or non-gradable courses were excluded from the dataset; 2) The features comprise categorical data, which is incompatible with machine learning algorithms. Consequently, it is necessary to transform these categorical course data into a numerical format before inputting them into the machine learning model. In this research, we employed a technique known as label encoding to convert our categorical data into numerical values \citep{patil2016categorical}. This method assigns an integer to each distinct nominal variable. Course subject and level features were transformed into integer numbers, representing the student's knowledge skill; 3) Courses sharing the same subject and level were treated as possessing the same knowledge skill. For example, courses like MATH 3201 and MATH 3800, both falling under the 3000-level math category, were considered part of the same skill set. Table \ref{tab_cp3} shows examples of the mapping between course subject and level encoded to integer number as $\text{skill\_id}$.

\begin{table}[!ht]
	\caption{ Examples of mapping between course subject and level encoded to integer number.}
\centering 
\begin{tabular}{ccc}
\hline
\textbf{Course Subject}  & \textbf{Course Level} & \textbf{Skill\_id} \\
 \hline
ACCT    & 2000   & 1                             \\ \hline
ACCT    & 3000   & 2                             \\ \hline
ACCT    & 4000   & 3                             \\ \hline
ACCT    & 5000   & 4                             \\ \hline
ADMN    & 5000   & 5                             \\ \hline
AFAM    & 1000   & 6                             \\ \hline
…       &        & …                             \\ \hline
…       &        & …                             \\ \hline
SPED    & 4000   & 231                           \\ \hline
SPED    & 5000   & 232                           \\ \hline
SPMT    & 1000   & 233                           \\ \hline
\end{tabular}
	\label{tab_cp3}
\end{table}

\subsection*{Statistical Analysis}
Over the four-year period, the total university (UNIV) dataset comprises $352,148$ records for undergraduate students, where Table \ref{tab_cp4} show the statistics of the dataset and sub datasets. 

\begin{table}[!ht]
	\caption{ Dataset statistics }
\centering 
\begin{tabular}{lccc}
\hline
    \textbf{Items}                          & \textbf{UNIV}   & \textbf{COE+COAS} & \textbf{COE}    \\ \hline
Total Records                 & 352,148 & 143,982  & 52,206 \\ \hline
Records after data cleaning   & 326,269 & 131,857  & 46,477 \\ \hline
Students                      & 17,181  & 7,210    & 2,387  \\ \hline
Types of Courses              & 2,124   & 1,797    & 1,027  \\ \hline
Types of Knowledge Components & 233     & 214      & 172    \\ \hline
\end{tabular}
	\label{tab_cp4}
\end{table}

Following the exclusion of incomplete, non-gradable, and no-grade information, $326,269$ records remain. A total of $17,181$ students participated in $2,124$ courses, with courses sharing the same subject and level treated as possessing the same knowledge skill. The research identifies a total of $233$ knowledge components (KCs) among all undergraduate students. In the subset of records from the College of Engineering (COE), the dataset comprises $52,206$ records. Following the exclusion of incomplete, non-gradable, and no-grade information, $46,477$ records remain. A total of $2,387$ students participated in $1,027$ courses. In the entirety of the university college system, the College of Arts and Sciences (COAS) stands as the largest academic entity. Consequently, we amalgamated student data from both COAS and COE disciplines to form another subset for training purposes. Within the subset of records originating from COE and COAS, the dataset encompasses $143,982$ entries. After removing incomplete, non-gradable, and no-grade data, $131,857$ records persist. Across $1,797$ courses, a collective of $7,210$ students took part in the study. This nuanced analysis not only provides insights into the overall predictive power of ML models but also delves into the domain-specific efficacy, showcasing the potential impact on targeted academic success initiatives. Table \ref{tab_cp5} represents the final dataset utilized for training and testing DKT models. The DKT models aim to discern correlations among courses, enabling the prediction of a student's likelihood of passing or failing a course based on historical records. 

\begin{table}[!ht]
	\caption{ Examples of samples for DKT at HBCUs. }
\centering 
\begin{tabular}{lccc}
\hline
\textbf{Index}  & \textbf{Universal ID} & \textbf{Skill\_id} & \textbf{Pass/Fail} \\ \hline
1      & 304883   & 167                           & 1                             \\ \hline
2      & 304883   & 200                           & 1                             \\ \hline
3      & 308438   & 82                            & 1                             \\ \hline
4      & 308438   & 1                             & 0                             \\ \hline
...    & …        & ...                           & ...                           \\ \hline
....   & …        & ...                           & ...                           \\ \hline
326267 & 6796570  & 66                            & 1                             \\ \hline
326268 & 6796693  & 2                             & 1                             \\ \hline
326269 & 6796693  & 149                           & 0                             \\ \hline
\end{tabular}
	\label{tab_cp5}
\end{table}

\section*{Methodology}
Deep Knowledge Tracing (DKT) has emerged as a powerful approach for modeling and predicting the knowledge mastery of learners in educational settings. Over time, several variants of DKT have been proposed, each introducing unique enhancements to address specific challenges. In this research, we delve into the methodologies of five prominent DKT models: DKT, DKT+, DKVMN, SAKT, and KQN to predict whether a student will pass or fail a course for HBUC education.
\begin{itemize}
\item Deep Knowledge Tracing (DKT) \citep{piech2015deep} is the first work to apply neural networks for knowledge tracing tasks. It uses recurrent neural network (RNNs) to effectively capture the temporal dependencies inherent within a sequence of interactions comprising a student's questions and corresponding answers. This enables the model to predict a student's response to a new question based on their historical interactions. 

A fundamental representation of a simple RNN network for DKT is defined as follows:
\begin{equation}
h_{t} = tanh(W_{hx}x_{t} + W_{hh}h_{t-1} + b_{h})
\end{equation}

\begin{equation}
y_{t} = \sigma(W_{yh}h_{t} +  b_{y})
\end{equation}
\smallskip

where \textit{tanh} is the activation function, \textit{$W_{hs}$} is the input weights, \textit{$W_{hh}$} is the recurrent weights, \textit{$W_{yt}$} is the readout weights, and \textit{$b_{h}$} and \textit{$b_{y}$} are the bias terms. 

Within the DKT framework, the hidden state $ht$ of the RNN is interpreted as the latent representation of the student's knowledge state. Moreover, \textit{h$_t$} is subjected to prediction through a Sigmoid-activated linear layer, denoted as \textit{y$_t$}. This layer is of the same length as the number of exercises, with each element representing the model's predicted probability of the student correctly answering the corresponding exercise. The empirical findings indicated that DKT surpassed traditional KT models across multiple benchmark datasets. This underscores the promise of employing deep learning models to tackle the KT challenge. Since then, deep learning-based methods reached state of the art on knowledge tracing \citep{wang2023dynamic}.
\item  DKT model with regularization (DKT+) \citep{yeung2018addressing} presents a notable extension of the DKT framework. It introduces regularization terms that correspond to reconstruction and waviness to the loss function of the original DKT model to enhance the consistency in prediction. The initial loss function is expanded by integrating three regularization terms, resulting in the subsequent regularized loss function: 
\begin{equation}
 \mathcal{L'} = \mathcal{L} + \lambda_{r}r + \lambda_{w1}w_1 + \lambda_{w2}w_2^2
\end{equation}
\smallskip

where \( \lambda_{r} \), \( \lambda_{w1} \), and \( \lambda_{w2} \) are regularization parameters.

 This refinement was devised to mitigate inherent constraints observed in DKT, particularly in its proficiency to accurately reconstruct input responses and diminish incongruities in predicting answers for questions associated with similar KCs. 
 
\item Dynamic Key-Value Memory Network (DKVMN) \citep{zhang2017dynamic} incorporates a key-value memory matrix, where each row represents an item or skill, and the columns correspond to key-value pairs associated with that item. The memory allows the model to store information about students' mastery of different skills and update this information as they interact with educational items. When a student responds to an item, the memory is updated to reflect their mastery of that item. The update process involves modifying the values associated with relevant keys in the memory, where the memory update for item $i$ and key $k$ is computed as:
\begin{equation}
\text{Memory}[i, k] = \text{Memory}[i, k] + \text{update}_k
\end{equation}
\smallskip
where $i$ is the index of the item, $k$ represents the key associated with the item, and $\text{update}_k$ is the update value for key $k$, computed based on the student's response to the item.

\item Self-Attentive Knowledge Tracing (SAKT) \citep{pandey2019selfattentive} tried to handles with sparse data which students interact with few KCs. It adopts a transformer-based architecture, replacing Long Short-Term Memory (LSTMs) with self-attention mechanisms to capture the relevance between the KCs and the students' historical interactions. SAKT employs a novel architecture that allows for the dynamic weighting of input sequences, enabling the model to focus on the most relevant interactions within a student's learning history. The self-attention mechanism computes attention scores for each pair of elements in a sequence and then calculates a weighted sum of the values based on these scores. Here's the formula for computing the attention score for element $i$ with respect to element $j$ is computed as:
\begin{equation}
\text{Attention}(Q, K, V)_{i,j} = \text{softmax}\left(\frac{Q_i \cdot K_j}{\sqrt{d_k}}\right) \cdot V_j
\end{equation}
\smallskip

where $Q$ is the query matrix, $K$ is the key matrix, $V$ is the value matrix, and $d_k$ is the dimensionality of the key vectors.
\vspace{\baselineskip}
\item Knowledge Query Network (KQN) \citep{lee2019knowledge} combines memory-augmented structures with attention mechanisms, encoding each student's current knowledge state as a query vector. The attention mechanisms enables KQN to retrieve pertinent information from its memory, prioritizing the most relevant items for each student. The attention weights are computed based on the similarity between each node and the current knowledge query, ensuring that the model prioritizes nodes most relevant to the student's current knowledge state. The attention weights $\alpha_{ij}$ for each node $j$ in the graph based on its similarity to the knowledge query $q_i$ are computed as:
\begin{equation}
\alpha_{ij} = \text{softmax}(q_i \cdot h_j)
\end{equation}
\smallskip

where $h_j$ represents the node embedding for node $j$, $q_i$ is the knowledge query for student $i$. Furthermore, KQN introduces a novel concept called probabilistic skill similarity, which relates pairwise cosine and Euclidean distances between skill vectors to the odds ratios of corresponding skills, making KQN interpretable and intuitive \citep{lee2019knowledge}.
\end{itemize}

\section*{Experiments}
It focuses on conducting a comparative evaluation of DKT models across different scales within the academic institution, specifically within the College of Engineering (COE) and at the university-wide level.

\subsection*{Dataset}
\begin{figure} [ht]
 	\centering
	\includegraphics[width=1.\linewidth]{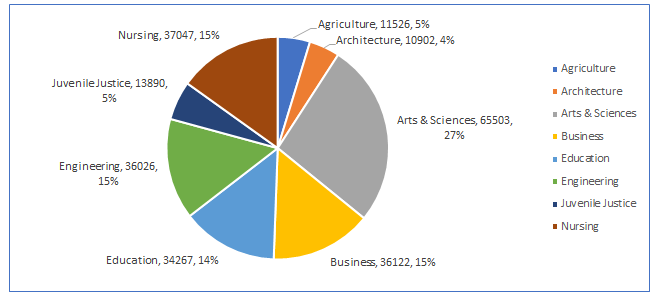}
	\caption{Training data distribution.}
	\label{Fig_chart1}
\end{figure}

We adhere to conventional practices by allocating dataset resources primarily for training purposes, typically dedicating 80\% to 90\% for this phase, while reserving the remaining portion for testing. Leveraging a four-year span of authentic data, we aim to predict student achievement in contemporary courses by analyzing historical academic performance. Specifically, the initial three years of data serve as the training dataset, while the fourth year is designated for testing. Figure \ref{Fig_chart1} offers an overview of the training data extracted from various colleges, with the College of Arts \& Sciences (COAS) emerging as the largest college within PVAMU. The training dataset encompasses three distinct subsets: data from the College of Engineering (COE), data from both the College of Engineering and the College of Arts and Sciences (COE + COAS), and data covering all undergraduate students across the university (UNIV). 

\begin{figure} [ht]
 	\centering
	\includegraphics[width=1.\linewidth]{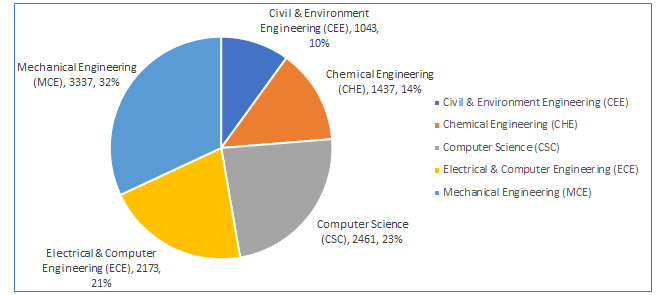}
	\caption{Testing data distribution.}
	\label{Fig_chart2}
\end{figure}

For the testing phase, we exclusively utilize data from the five departments within the College of Engineering: Civil and Environment Engineering (CEE), Chemical Engineering (CHE), Computer Science (CSC), Electrical and Computer Engineering (ECE), and Mechanical Engineering (MCE). Detailed insights into these Engineering departments are depicted in Figure \ref{Fig_chart2}.

Furthermore, Table~\ref{tab_cp60} presents details of data distributions for training and testing of DKT.

\begin{table}[!ht]
	\caption{ Dataset statistics }
\centering 
\begin{tabular}{lrrr}
\hline
Training   Dataset & \# of records & \# of students & \# of KCs \\ \hline
COE                & 36,026        & 2,036          & 165       \\
COE+COAS           & 101,529       & 6,179          & 206       \\
UNIV               & 246,964       & 14,549         & 224       \\ \hline
Testing Dataset & \# of records & \# of students & \# of KCs        \\ \hline
CEE                & 1,043         & 130            & 57        \\
CHE                & 1,437         & 147            & 57        \\
CSC                & 2,461         & 284            & 67        \\
ECE                & 2,173         & 244            & 75        \\
MCE                & 3,337         & 387            & 83        \\
COE                & 10,451        & 1,182          & 125       \\
UNIV               & 79,305        & 9,102          & 200       \\ \hline
\end{tabular}
	\label{tab_cp60}
\end{table}

\subsection*{Experiment Setup}

In this study, we employed the default settings and configurations of DKT models sourced from the GitHub repository. The original PyTorch implementation of these DKT models can be accessed at https://github.com/hcnoh/knowledge-tracing-collection-pytorch. 
\begin{itemize}
\item 
Batch Size: The batch size denotes the number of instances processed in a single iteration during the training process. The default value is set to 256.
\item 
Number of Epochs: Epochs represent the number of complete passes through the entire training dataset. The default number of epochs utilized in this research is 100.
\item 
Learning Rate: This parameter governs the step size during the optimization process of the training algorithm. The default learning rate employed in this study is 0.001.
\item 
Optimizer: The optimizer determines the specific algorithm employed for optimizing the model parameters during the training process. The available optimizers include Stochastic Gradient Descent (SGD) and Adaptive Moment Estimation (Adam). For this research, the default optimizer chosen is Adam.
\item 
Sequence Length: The sequence length represents the number of time steps or elements considered in the dataset for each training instance. The default sequence length utilized in this research is 100.
\end{itemize}

These default settings serve as the foundational configuration for the DKT models applied throughout the experimental investigations conducted in this study. These DKT models were developed using PyTorch and trained on a workstation equipped with NVIDIA V-100 GPUs.

\subsection*{Evaluation Metrics}

In the realm of knowledge tracing, numerous methodologies employ binary classification techniques to forecast students' academic performance, such as determining the accuracy of exercise completion \citep{song2022survey}. This study aims to evaluate the effectiveness of various DKT models in predicting students' likelihood of success in a course, leveraging their historical subject grades as predictive features. All DKT models are subjected to a binary classification task that is to predict whether a student will pass or fail the course. The metrics used are accuracy, recall, precision, F1-score, and the Area Under the Receiver Operating Characteristic Curve (AUC). 

\begin{equation}
	Accuracy = \frac{TP+TN}{TP+FP+TN+FN}
\end{equation}
\smallskip

$Accuracy$, a commonly employed metric in classification tasks, measures the ratio of correctly classified instances, encompassing both true positives and true negatives, relative to the total instances evaluated, thereby providing a straightforward assessment of prediction accuracy.

\begin{equation}
	Precision = \frac{TP}{TP+FP}
\end{equation}

\begin{equation}
	Recall = \frac{TP}{TP+FN}
\end{equation}

\begin{equation}
	F1 = 2* \frac{Precision *Recall}{Precision+Recall}
\end{equation}

where True Positive (TP), False Positive (FP), True Negative (TN), and False Negative (FN) come from confusion matrix terminology.

\begin{itemize}
\item TP: correct predictions of passing courses.
\item TN: correct predictions of failing courses.
\item FP: incorrect predictions of passing courses.
\item FN: incorrect predictions of failing courses.
\end{itemize}

$Precision$ measures the proportion of true positive predictions out of all positive predictions. $Recall$ measures the proportion of true positive predictions out of all actual positive instances. The $F1$ score is a metric used to evaluate the performance of a classification model, particularly when dealing with imbalanced classes. It combines both precision and recall into a single measure.

Conversely, AUC serves as a valuable metric for assessing the performance of binary classifiers, particularly in scenarios characterized by class imbalance or variable importance of false positives and false negatives. A higher AUC value signifies superior discrimination capabilities of the model in distinguishing between positive and negative classes~\citep{gervet2020when}. 

\subsection*{Results and Discussion}

The findings and discussions present the comprehensive prediction performance analysis of DKT models across various testing and training datasets, as delineated in Tables \ref{tab_ex1auc} through \ref{tab_ex45f1}. These tables encapsulate the AUC and accuracy scores, providing insights into the model's efficacy. Recall, Precision, and F1 scores, further enriching our understanding of the model's capabilities under different training and testing conditions.

\subsubsection*{Knowledge tracing on departments within the COE when training on COE} Table \ref{tab_ex1auc} and Table \ref{tab_ex1f1} illustrate the performance of these models when tested on five engineering departments with training on the COE dataset. DKT and KQN models demonstrate superior performance compared to other models. Specifically, the DKT model achieves an average AUC of 0.6267, accuracy of 0.7604, and F1 score of 0.8567. The KQN model performs similarly well, with an AUC of 0.6255, accuracy of 0.7817, and F1 score of 0.8692. In contrast, the DKVMN and SAKT models exhibit moderate performance. The DKVMN model records an AUC of 0.5921, accuracy of 0.7892, and F1 score of 0.8779, while the SAKT model shows an AUC of 0.6099, accuracy of 0.7668, and F1 score of 0.8590. Among these models, the DKT+ model performs the least effectively, with an AUC of 0.5461, accuracy of 0.6277, and F1 score of 0.7469.

\subsubsection*{Knowledge tracing on departments within the COE when training on COE and COAS} 
Similarly, Table \ref{tab_ex2auc} and Table \ref{tab_ex2f1}, testing the same engineering departments but with training on the COE + COAS dataset, indicates that the KQN models continue to deliver strong performance, with average metrics of AUC 0.6820, accuracy 0.8092, and F1 score 0.8884. The SAKT and DKVMN models show noticeable improvements compared to previous experiments and have surpassed the DKT model. Specifically, the SAKT model achieves an AUC of 0.6692, accuracy of 0.7903, and F1 score of 0.8750, while the DKVMN model reports an AUC of 0.6618, accuracy of 0.7949, and F1 score of 0.8805. In comparison, the DKT model has an average AUC of 0.6474, accuracy of 0.7897, and F1 score of 0.8768.

\begin{landscape}
\begin{table*} [ht]
\scriptsize
\centering 
\caption{ Performance comparison between engineering departments via Accuracy and AUC when DKT models are trained on on College of Engineering (COE) data. EM denotes evaluation metrics. }

\begin{tabular}{lcccccccccc}
\toprule
Model & \multicolumn{2}{c}{DKT} & \multicolumn{2}{c}{DKT+} & \multicolumn{2}{c}{DKVMN} & \multicolumn{2}{c}{SAKT} & \multicolumn{2}{c}{KQN} \\
\toprule
EM & AUC & Accuracy & AUC & Accuracy & AUC & Accuracy & AUC & Accuracy & AUC & Accuracy \\
\midrule
CEE & 0.6030 & 0.6840 & 0.5302 & 0.5828 & 0.5690 & 0.7288 & 0.5787 & 0.7131 & 0.5523 & 0.6805 \\
CHE & 0.6605 & 0.8029 & 0.5583 & 0.6419 & 0.6069 & 0.8509 & 0.6851 & 0.8270 & 0.6725 & 0.8486 \\
CSC & 0.5963 & 0.8011 & 0.5370 & 0.6658 & 0.5262 & 0.8378 & 0.6327 & 0.8132 & 0.5816 & 0.8179 \\
ECE & 0.6169 & 0.7655 & 0.5484 & 0.6481 & 0.6202 & 0.7875 & 0.5758 & 0.7872 & 0.6531 & 0.7922 \\
MCE & 0.6568 & 0.7487 & 0.5582 & 0.5998 & 0.6384 & 0.7412 & 0.5772 & 0.6934 & 0.6678 & 0.7695 \\
\midrule
Average & 0.6267 & 0.7604 & 0.5464 & 0.6277 & 0.5921 & 0.7892 & 0.6099 & 0.7668 & 0.6255 & 0.7817 \\
\bottomrule
\end{tabular}
\label{tab_ex1auc}
\end{table*}

\begin{table*}[ht]
\centering 
\caption{ Performance comparison between engineering departments via Recall, Precision and F1 scores between engineering departments when DKT models are trained on College of Engineering (COE) data. }

\tiny

\begin{tabular}{lccccccccccccccc}
\toprule
Model   & \multicolumn{3}{c}{DKT}     & \multicolumn{3}{c}{DKT+}    & \multicolumn{3}{c}{DKVMN}   & \multicolumn{3}{c}{SAKT}    & \multicolumn{3}{c}{KQN}     \\
\toprule
        EM & Recall & Precision & F1     & Recall & Precision & F1     & Recall & Precision & F1     & Recall & Precision & F1     & Recall & Precision & F1     \\
\midrule
CEE     & 0.8496 & 0.7536    & 0.7987 & 0.6427 & 0.7548    & 0.6943 & 0.9101 & 0.7719    & 0.8353 & 0.8959 & 0.7588    & 0.8217 & 0.8322 & 0.7583    & 0.7935 \\
CHE     & 0.8923 & 0.8794    & 0.8858 & 0.6891 & 0.8658    & 0.7674 & 0.9631 & 0.8769    & 0.9180 & 0.9038 & 0.8958    & 0.8998 & 0.9726 & 0.8672    & 0.9169 \\
CSC     & 0.9304 & 0.8472    & 0.8869 & 0.7285 & 0.8551    & 0.7867 & 0.9752 & 0.8546    & 0.9109 & 0.9311 & 0.8591    & 0.8937 & 0.9421 & 0.8568    & 0.8974 \\
ECE     & 0.9194 & 0.8123    & 0.8625 & 0.7267 & 0.8132    & 0.7675 & 0.9620 & 0.8091    & 0.8790 & 0.9477 & 0.8161    & 0.8770 & 0.9300 & 0.8307    & 0.8775 \\
MCE     & 0.9236 & 0.7869    & 0.8498 & 0.6638 & 0.7830    & 0.7185 & 0.9314 & 0.7752    & 0.8462 & 0.8111 & 0.7948    & 0.8029 & 0.9256 & 0.8045    & 0.8608 \\
\midrule
Average & 0.9031 & 0.8159    & 0.8567 & 0.6902 & 0.8144    & 0.7469 & 0.9484 & 0.8175    & 0.8779 & 0.8979 & 0.8249    & 0.8590 & 0.9205 & 0.8235    & 0.8692 \\
\bottomrule
\end{tabular}
	\label{tab_ex1f1}
\end{table*}
\end{landscape}

\begin{landscape}
\begin{table*}[ht]
\scriptsize
\centering 
\caption{Comparing AUC and Accuracy for each engineering department when DKT models are  trained on COE and COAS data. }

\begin{tabular}{lcccccccccc}
\toprule
Model & \multicolumn{2}{c}{DKT} & \multicolumn{2}{c}{DKT+} & \multicolumn{2}{c}{DKVMN} & \multicolumn{2}{c}{SAKT} & \multicolumn{2}{c}{KQN} \\
\toprule
EM & AUC & Accuracy & AUC & Accuracy & AUC & Accuracy & AUC & Accuracy & AUC & Accuracy \\
\midrule

CEE     & 0.5578    & 0.7046      & 0.4841     & 0.5753      & 0.6106     & 0.7251       & 0.6015     & 0.7183      & 0.6187    & 0.7337      \\
CHE     & 0.6974    & 0.8593      & 0.6627     & 0.7171      & 0.7342     & 0.8431       & 0.6992     & 0.8715      & 0.7162    & 0.8819      \\
CSC     & 0.7012    & 0.8382      & 0.6107     & 0.7261      & 0.6935     & 0.8533       & 0.6653     & 0.8021      & 0.7017    & 0.8433      \\
ECE     & 0.6324    & 0.7930      & 0.6314     & 0.7126      & 0.6455     & 0.7954       & 0.6978     & 0.7931      & 0.6713    & 0.8065      \\
MCE     & 0.6480    & 0.7536      & 0.5745     & 0.6707      & 0.6254     & 0.7574       & 0.6823     & 0.7666      & 0.7022    & 0.7804      \\
\midrule
Average & 0.6474    & 0.7897      & 0.5927     & 0.6804      & 0.6618     & 0.7949       & 0.6692     & 0.7903      & 0.6820    & 0.8092      \\
\bottomrule
\end{tabular}
\label{tab_ex2auc}
\end{table*}

\begin{table*}[ht]
\centering 
\caption{ Comparing Recall, Precision and F1 scores for each engineering department when DKT models are  trained on COE and COAS data. }

\tiny

\begin{tabular}{lccccccccccccccc}
\toprule
Model   & \multicolumn{3}{c}{DKT}     & \multicolumn{3}{c}{DKT+}    & \multicolumn{3}{c}{DKVMN}   & \multicolumn{3}{c}{SAKT}    & \multicolumn{3}{c}{KQN}     \\
\toprule
        EM & Recall & Precision & F1     & Recall & Precision & F1     & Recall & Precision & F1     & Recall & Precision & F1     & Recall & Precision & F1     \\
\midrule

CEE     & 0.9188 & 0.7419    & 0.8209 & 0.6722 & 0.7302    & 0.7000 & 0.9098 & 0.7691    & 0.8336 & 0.9167 & 0.7543    & 0.8276 & 0.9455 & 0.7552    & 0.8397 \\
CHE     & 0.9550 & 0.8891    & 0.9209 & 0.7644 & 0.8903    & 0.8226 & 0.9249 & 0.8977    & 0.9111 & 0.9637 & 0.8948    & 0.9280 & 0.9758 & 0.8959    & 0.9341 \\
CSC     & 0.9419 & 0.8758    & 0.9076 & 0.8017 & 0.8630    & 0.8312 & 0.9701 & 0.8713    & 0.9180 & 0.8774 & 0.8873    & 0.8823 & 0.9365 & 0.8844    & 0.9097 \\
ECE     & 0.9418 & 0.8244    & 0.8792 & 0.8103 & 0.8269    & 0.8185 & 0.9632 & 0.8152    & 0.8830 & 0.9244 & 0.8350    & 0.8774 & 0.9566 & 0.8284    & 0.8879 \\
MCE     & 0.9478 & 0.7794    & 0.8554 & 0.7811 & 0.7885    & 0.7848 & 0.9474 & 0.7818    & 0.8567 & 0.9304 & 0.7993    & 0.8599 & 0.9604 & 0.7964    & 0.8707 \\ 
\midrule
Average & 0.9411 & 0.8221    & 0.8768 & 0.7659 & 0.8198    & 0.7914 & 0.9431 & 0.8270    & 0.8805 & 0.9225 & 0.8341    & 0.8750 & 0.9550 & 0.8321    & 0.8884  \\ 
\bottomrule
\end{tabular}
	\label{tab_ex2f1}
\end{table*}
\end{landscape}

\subsubsection*{Knowledge tracing on departments within the COE when training on UNIV data} In Table \ref{tab_ex3auc} and Table \ref{tab_ex3f1}, where testing is conducted on the same engineering departments but with training on the UNIV dataset, all models exhibit moderate to strong performance. The analysis reveals that the KQN model consistently delivers superior performance, with an average AUC of 0.6965, accuracy of 0.7996, and F1 score of 0.8834. The SAKT and DKVMN models also show competitive performance, reflecting their adaptability across various domains. Specifically, the SAKT model averages an AUC of 0.6764, accuracy of 0.8028, and F1 score of 0.8831, while the DKVMN model achieves an AUC of 0.6676, accuracy of 0.8066, and F1 score of 0.8889.
Overall, the KQN model stands out for its robust performance across different testing and training datasets, establishing it as a reliable choice for educational prediction tasks. The SAKT and DKVMN models also demonstrate strong performance, suggesting their suitability for real-world educational prediction applications alongside KQN. In contrast, the DKT and DKT+ models show moderate performance and are generally surpassed by the KQN and SAKT models, underscoring the advantages of newer model architectures.

\subsubsection*{Knowledge tracing on COE and UNIV when training on UNIV data} Table \ref{tab_ex45auc} and Table \ref{tab_ex45f1} illustrate the performance of these models when tested on COE and UNIV with training on the UNIV dataset. For COE testing dataset, the analysis shows that the KQN model exhibits strong performance across datasets, with an average AUC of 0.6847, accuracy of 0.7963, and F1 score of 0.8805. The SAKT model also performs competitively, with an average AUC of 0.6845, accuracy of 0.7848, and F1 score of 0.8712. In contrast, the DKVMN and DKT models display moderate performance. The DKVMN model has an AUC of 0.6525, accuracy of 0.7987, and F1 score of 0.8837, while the DKT model shows an AUC of 0.6579, accuracy of 0.7779, and F1 score of 0.8672. 

For the UNIV testing dataset, both the SAKT and KQN models maintain strong performance. The SAKT model averages an AUC of 0.6644, accuracy of 0.8094, and F1 score of 0.8892, while the KQN model achieves an AUC of 0.6639, accuracy of 0.8197, and F1 score of 0.8974. The DKVMN and DKT models continue to show moderate performance, with the DKVMN model having an AUC of 0.6326, accuracy of 0.8153, and F1 score of 0.8961, and the DKT model reporting an AUC of 0.6500, accuracy of 0.7917, and F1 score of 0.8791.

When evaluating the models trained on the UNIV dataset, it is observed that testing on the smaller COE dataset yields better performance metrics than testing on the entire UNIV dataset. This discrepancy is largely due to the differences in testing and training ratios. The smaller COE testing dataset might have fewer examples but could present a more homogeneous and controlled environment, allowing the model to perform more effectively. In contrast, the larger UNIV dataset introduces greater variability and complexity, which can challenge the model's ability to generalize, potentially leading to lower performance metrics.

\begin{landscape}
\begin{table*}[ht]
\scriptsize
\centering 
\caption{ Comparing AUC and Accuracy for each engineering department when DKT models are  trained on university (UNIV) data. }

\begin{tabular}{lcccccccccc}
\toprule
Model & \multicolumn{2}{c}{DKT} & \multicolumn{2}{c}{DKT+} & \multicolumn{2}{c}{DKVMN} & \multicolumn{2}{c}{SAKT} & \multicolumn{2}{c}{KQN} \\
\toprule
EM & AUC & Accuracy & AUC & Accuracy & AUC & Accuracy & AUC & Accuracy & AUC & Accuracy \\
\midrule
CEE     & 0.6259    & 0.7213      & 0.5031     & 0.5843      & 
0.6278     & 0.7535       & 0.6584     & 0.7271      & 0.6468    & 0.7308      \\
CHE     & 0.7109    & 0.8508      & 0.6505     & 0.6953      & 0.7097     & 0.8450       & 0.7129     & 0.8653      & 0.7221    & 0.8536      \\
CSC     & 0.6930    & 0.8336      & 0.6658     & 0.7298      & 0.6963     & 0.8637       & 0.6875     & 0.8584      & 0.7475    & 0.8492      \\
ECE     & 0.6115    & 0.7309      & 0.5945     & 0.6737      & 0.6463     & 0.8095       & 0.6332     & 0.7935      & 0.6720    & 0.7915      \\
MCE     & 0.6881    & 0.7598      & 0.5981     & 0.6950      & 0.6578     & 0.7614       & 0.6899     & 0.7697      & 0.6943    & 0.7731      \\ 
\midrule
Average & 0.6659    & 0.7793      & 0.6024     & 0.6756      & 0.6676     & 0.8066       & 0.6764     & 0.8028      & 0.6965    & 0.7996     \\ 
\bottomrule
\end{tabular}
	\label{tab_ex3auc}

\end{table*}

\begin{table*}[ht]
\centering 
\caption{ Comparing Recall, Precision and F1 scores for each engineering department when DKT models are  trained on university (UNIV) data. }

\tiny

\begin{tabular}{lccccccccccccccc}
\toprule
Model   & \multicolumn{3}{c}{DKT}     & \multicolumn{3}{c}{DKT+}    & \multicolumn{3}{c}{DKVMN}   & \multicolumn{3}{c}{SAKT}    & \multicolumn{3}{c}{KQN}     \\
\toprule
        EM & Recall & Precision & F1     & Recall & Precision & F1     & Recall & Precision & F1     & Recall & Precision & F1     & Recall & Precision & F1     \\
\midrule
        EM & Recall & Precision & F1     & Recall & Precision & F1     & Recall & Precision & F1     & Recall & Precision & F1     & Recall & Precision & F1     \\ \hline
CEE     & 0.9192 & 0.7557    & 0.8295 & 0.6961 & 0.7275    & 0.7115 & 0.9617 & 0.7697    & 0.8551 & 0.9156 & 0.7622    & 0.8319 & 0.9362 & 0.7565    & 0.8368 \\
CHE     & 0.9604 & 0.8773    & 0.9170 & 0.7484 & 0.8785    & 0.8082 & 0.9417 & 0.8867    & 0.9134 & 0.9596 & 0.8918    & 0.9245 & 0.9645 & 0.8772    & 0.9188 \\
CSC     & 0.9403 & 0.8723    & 0.9050 & 0.8059 & 0.8641    & 0.8340 & 0.9797 & 0.8747    & 0.9242 & 0.9542 & 0.8871    & 0.9194 & 0.9859 & 0.8566    & 0.9167 \\
ECE     & 0.8495 & 0.8205    & 0.8347 & 0.7579 & 0.8207    & 0.7881 & 0.9806 & 0.8181    & 0.8920 & 0.9482 & 0.8214    & 0.8803 & 0.9447 & 0.8216    & 0.8789 \\
MCE     & 0.9090 & 0.8042    & 0.8534 & 0.8163 & 0.7936    & 0.8048 & 0.9585 & 0.7799    & 0.8600 & 0.9154 & 0.8100    & 0.8595 & 0.9516 & 0.7941    & 0.8657 \\ \midrule
Average & 0.9157 & 0.8260    & 0.8679 & 0.7649 & 0.8169    & 0.7893 & 0.9644 & 0.8258    & 0.8889 & 0.9386 & 0.8345    & 0.8831 & 0.9566 & 0.8212    & 0.8834 \\ \bottomrule
\end{tabular}
	\label{tab_ex3f1}

\end{table*}
\end{landscape}

\begin{landscape}
\begin{table*}[ht]
\scriptsize
\centering 
\caption{Comparing AUC and Accuracy for COE and UNIV when DKT models are  trained on university (UNIV) data.}

\begin{tabular}{lcccccccccc}
\toprule
Model & \multicolumn{2}{c}{DKT} & \multicolumn{2}{c}{DKT+} & \multicolumn{2}{c}{DKVMN} & \multicolumn{2}{c}{SAKT} & \multicolumn{2}{c}{KQN} \\
\toprule
EM & AUC & Accuracy & AUC & Accuracy & AUC & Accuracy & AUC & Accuracy & AUC & Accuracy \\
\midrule
COE   & 0.6579    & 0.7779      & 0.5805     & 0.6551      & 0.6525     & 0.7987       & 0.6845     & 0.7848      & 0.6847    & 0.7963      \\
UNIV  & 0.6500    & 0.7917      & 0.5773     & 0.6700      & 0.6326     & 0.8153       & 0.6644     & 0.8094      & 0.6639    & 0.8197      \\ \midrule
\end{tabular}

	\label{tab_ex45auc}
\end{table*}

\begin{table*}[ht]
\centering
\caption{ Comparing Recall, Precision and F1 scores for COE and UNIV when DKT models are  trained on university (UNIV) data. }
\tiny

\begin{tabular}{lccccccccccccccc}
\toprule
Model   & \multicolumn{3}{c}{DKT}     & \multicolumn{3}{c}{DKT+}    & \multicolumn{3}{c}{DKVMN}   & \multicolumn{3}{c}{SAKT}    & \multicolumn{3}{c}{KQN}     \\
\toprule
        EM & Recall & Precision & F1     & Recall & Precision & F1     & Recall & Precision & F1     & Recall & Precision & F1     & Recall & Precision & F1     \\
\midrule

COE   & 0.9040 & 0.8333    & 0.8672 & 0.7275 & 0.8220    & 0.7719 & 0.9489 & 0.8269    & 0.8837 & 0.9079 & 0.8374    & 0.8712 & 0.9351 & 0.8319    & 0.8805 \\
UNIV  & 0.9132 & 0.8474    & 0.8791 & 0.7360 & 0.8460    & 0.7872 & 0.9642 & 0.8369    & 0.8961 & 0.9224 & 0.8583    & 0.8892 & 0.9516 & 0.8491    & 0.8974 \\ 
\midrule
\end{tabular}
	\label{tab_ex45f1}
\end{table*}
\end{landscape}

Both SAKT and KQN exhibit superior performance compared to traditional other DKT models due to their innovative architectural choices. Unlike DKT, which relies on RNNs or LSTMs networks to capture temporal dependencies, SAKT and KQN leverage more advanced techniques. SAKT employs transformer-based architectures, which can capture long-range dependencies in sequential data using self-attention mechanisms, while KQN utilizes query mechanisms to represent interactions between students and items, capturing complex dependencies more effectively than traditional sequential models. SAKT utilizes attention mechanisms, while KQN employs the dot product process to dynamically focus on relevant information when making predictions. Both models adaptively select the most informative items for each student, enhancing their ability to tailor learning experiences and improve educational outcomes. These approaches allow SAKT and KQN to better adapt to the nuances of student learning patterns, resulting in enhanced accuracy and efficiency in knowledge tracing tasks.

Moreover, as evident from the provided dataset statistics Table \ref{tab_cp60} and the information gleaned from Tables \ref{tab_ex1auc}-\ref{tab_ex45f1} , there is a discernible trend indicating that the prediction performance of DKT models tends to improve as the training dataset increases in size and diversity. This trend is particularly notable when comparing performance across different training datasets. For instance, transitioning from smaller training datasets like COE, with 36,026 records, to larger and more comprehensive ones like UNIV, with 246,964 records, results in a notable improvement in prediction performance metrics (AUC, accuracy, and F1) across different testing datasets. 

Additionally, combining training datasets, as observed in the COE+COAS scenario, leads to a more diverse and extensive dataset (101,529 records), further enhancing model performance. The increase in training data size enables the models to learn more effectively from a broader range of student interactions and knowledge components, thereby leading to improved prediction accuracy. Overall, the observed trend suggests that scaling up the training dataset size and diversity positively impacts the prediction performance of DKT models. This underscores the importance of utilizing large and diverse datasets in educational data mining tasks, as it enables more accurate predictions of student knowledge mastery and learning outcomes.

\section*{Related Work}
Knowledge tracing involves modeling students' learning progress based on their activity sequences, making it a challenging task due to the need for accurate performance prediction and understanding of students' mastery levels. Researchers have tackled this challenge through various approaches. There are two most representative models: Bayesian Knowledge Tracking (BKT) and Deep Knowledge Tracing (DKT). Early efforts focused on BKT models \citep{corbett1994knowledge}, wherein each student's knowledge state is represented as a binary variable. These models employ probabilistic techniques like the Hidden Markov Model (HMM) to assess students' grasp of concepts \citep{awad2015hidden}. It characterizes a student's progress in tackling problems associated with a specific concept by utilizing a binary indicator (either correct or incorrect) and consistently refines its assessment of the student's understanding of that concept \citep{mao2018deep}.

DKT \citep{piech2015deep}, a pioneering algorithm designed to model students’ learning states through recurrent neural networks (RNN), has demonstrated significant enhancements in predictive accuracy. It grasps the temporal nuances within a sequence of student-question interactions. Empirical findings demonstrated DKT's superiority over traditional knowledge tracing models across various benchmark datasets \citep{abdelrahman2023knowledge}. The DKT models can be categorized into the following groups~\citep{song2022survey}:

\begin{itemize}
\item DKT and its Variants. Early deep learning KT models often used RNNs architectures like Long Short-Term Memory (LSTM) or Gated Recurrent Unit (GRU). RNNs are well-suited for sequential data, making them appropriate for modeling students' learning trajectories over time. e.g., DKT \citep{piech2015deep} and DKT+ \citep{yeung2018addressing}.

\item Memory Network based models. Memory networks introduce an external memory component that the model can read from and write to at each time step. This memory serves as a repository for storing past interactions, knowledge concepts, or context information. By accessing this memory, the model can maintain a richer representation of the student's learning history and better infer their current knowledge state. e.g., DKVMN \citep{zhang2017dynamic}, and SKVMN \citep{Abdelrahman_2019}.

\item Attention Mechanism based models. To improve the modeling of long-term dependencies and focus on relevant parts of the input sequence, attention mechanisms were introduced in KT models. Attention mechanisms allow the model to dynamically weigh the importance of different time steps or features, enhancing its predictive power. e.g., SAKT \citep{pandey2019selfattentive}, DKVMN \citep{zhang2017dynamic} and SAINT \citep{choi2020towards}.

\item Graph Structure based models. Another direction of development involves using graph neural networks to represent the knowledge structure and student interactions as graphs. Graph Neural Networks (GNNs) allow models to capture complex relationships between knowledge concepts and how they influence each other's mastery. e.g., GKT \citep{nakagawa2019graph}.

\end{itemize}

The pioneering use of DKT in online exercises has significantly advanced the prediction of students' performance. By leveraging past student records, the original DKT model successfully forecasts whether a student will pass or fail their current academic exercises. This research extends beyond mere prediction, utilizing past grades to anticipate students' future course outcomes. The research findings underscore the effectiveness of DKT models in accurately predicting students' academic trajectories, thereby highlighting their potential for enhancing educational outcomes.

\section*{Conclusion and Future Work}
PAL allows for monitoring individual students' progress and tailoring their learning paths to their unique knowledge and needs. DKT for PAL  enhanced modeling students' evolving knowledge to predict their future performance. This study mitigates the gap in its investigation and application within specific educational contexts, particularly at HBCUs through building a comprehensive dataset from PVAMU and DKT model evaluation on this dataset. This study advocates for the utilization of DKT models to forecast the academic performance of undergraduate students, particularly in determining whether they will pass or fail a course based on their historical grade records. Through the utilization of diverse training datasets and various DKT models, the findings affirm the efficacy of DKT models in predicting students' academic outcomes for HBCUs. Furthermore, the adoption of DKT models in academic settings can serve as a valuable tool for enhancing student support mechanisms and promoting academic success for HBCUs.

In the future, it plans to explore the potential of adopting different DKT architectures not considered in this study such as large language models (LLMs)  to develop more sophisticated models capable of capturing complex patterns in student learning data. Furthermore, there is a need to expand the way of knowledge skills are determined within the DKT models. While current research rely on course subject and course level, future research could explore incorporating additional course features to provide a more comprehensive understanding of student knowledge and learning trajectories. By incorporating these advancements, future DKT models have the potential to offer more accurate and personalized insights into student learning processes, ultimately leading to improved educational outcomes.

\bmhead{Acknowledgements}
This research work is supported by the U.S. NSF awards 2205891 and 2235731. The U.S. Government is authorized to reproduce and distribute reprints for governmental purposes notwithstanding any copyright notation thereon. The views and conclusions contained herein are those of the authors and should not be interpreted as necessarily representing the official policies or endorsements, either expressed or implied, of the U.S. NSF or the U.S. Government.

\bibliography{sn-bibliography}


\begin{thebibliography}{23}
\ifx \bisbn   \undefined \def \bisbn  #1{ISBN #1}\fi
\ifx \binits  \undefined \def \binits#1{#1}\fi
\ifx \bauthor  \undefined \def \bauthor#1{#1}\fi
\ifx \batitle  \undefined \def \batitle#1{#1}\fi
\ifx \bjtitle  \undefined \def \bjtitle#1{#1}\fi
\ifx \bvolume  \undefined \def \bvolume#1{\textbf{#1}}\fi
\ifx \byear  \undefined \def \byear#1{#1}\fi
\ifx \bissue  \undefined \def \bissue#1{#1}\fi
\ifx \bfpage  \undefined \def \bfpage#1{#1}\fi
\ifx \blpage  \undefined \def \blpage #1{#1}\fi
\ifx \burl  \undefined \def \burl#1{\textsf{#1}}\fi
\ifx \doiurl  \undefined \def \doiurl#1{\url{https://doi.org/#1}}\fi
\ifx \betal  \undefined \def \betal{\textit{et al.}}\fi
\ifx \binstitute  \undefined \def \binstitute#1{#1}\fi
\ifx \binstitutionaled  \undefined \def \binstitutionaled#1{#1}\fi
\ifx \bctitle  \undefined \def \bctitle#1{#1}\fi
\ifx \beditor  \undefined \def \beditor#1{#1}\fi
\ifx \bpublisher  \undefined \def \bpublisher#1{#1}\fi
\ifx \bbtitle  \undefined \def \bbtitle#1{#1}\fi
\ifx \bedition  \undefined \def \bedition#1{#1}\fi
\ifx \bseriesno  \undefined \def \bseriesno#1{#1}\fi
\ifx \blocation  \undefined \def \blocation#1{#1}\fi
\ifx \bsertitle  \undefined \def \bsertitle#1{#1}\fi
\ifx \bsnm \undefined \def \bsnm#1{#1}\fi
\ifx \bsuffix \undefined \def \bsuffix#1{#1}\fi
\ifx \bparticle \undefined \def \bparticle#1{#1}\fi
\ifx \barticle \undefined \def \barticle#1{#1}\fi
\bibcommenthead
\ifx \bconfdate \undefined \def \bconfdate #1{#1}\fi
\ifx \botherref \undefined \def \botherref #1{#1}\fi
\ifx \url \undefined \def \url#1{\textsf{#1}}\fi
\ifx \bchapter \undefined \def \bchapter#1{#1}\fi
\ifx \bbook \undefined \def \bbook#1{#1}\fi
\ifx \bcomment \undefined \def \bcomment#1{#1}\fi
\ifx \oauthor \undefined \def \oauthor#1{#1}\fi
\ifx \citeauthoryear \undefined \def \citeauthoryear#1{#1}\fi
\ifx \endbibitem  \undefined \def \endbibitem {}\fi
\ifx \bconflocation  \undefined \def \bconflocation#1{#1}\fi
\ifx \arxivurl  \undefined \def \arxivurl#1{\textsf{#1}}\fi
\csname PreBibitemsHook\endcsname

\bibitem[\protect\citeauthoryear{Awad and Khanna}{2015}]{awad2015hidden}
\begin{bchapter}
\bauthor{\bsnm{Awad}, \binits{M.}},
\bauthor{\bsnm{Khanna}, \binits{R.}}:
\bctitle{Hidden markov model}.
In: \bbtitle{Efficient Learning Machines}.
\bpublisher{Apress},
\blocation{Berkeley, CA}
(\byear{2015})
\end{bchapter}
\endbibitem

\bibitem[\protect\citeauthoryear{Abdelrahman and Wang}{2019}]{Abdelrahman_2019}
\begin{barticle}
\bauthor{\bsnm{Abdelrahman}, \binits{G.}},
\bauthor{\bsnm{Wang}, \binits{Q.}}:
\batitle{Knowledge tracing with sequential key-value memory networks}.
\bjtitle{Proceedings of the ACM on Human-Computer Interaction}
(\byear{2019})
\doiurl{10.1145/3331184.3331195}
\end{barticle}
\endbibitem

\bibitem[\protect\citeauthoryear{Abdelrahman
  et~al.}{2023}]{abdelrahman2023knowledge}
\begin{barticle}
\bauthor{\bsnm{Abdelrahman}, \binits{G.}},
\bauthor{\bsnm{Wang}, \binits{Q.}},
\bauthor{\bsnm{Nunes}, \binits{B.}}:
\batitle{Knowledge tracing: A survey}.
\bjtitle{ACM Computing Surveys}
\bvolume{55}(\bissue{11}),
\bfpage{1}--\blpage{37}
(\byear{2023})
\doiurl{10.1145/3569576}
\end{barticle}
\endbibitem

\bibitem[\protect\citeauthoryear{Bajaj and Sharma}{2018}]{bajaj2018smart}
\begin{barticle}
\bauthor{\bsnm{Bajaj}, \binits{R.}},
\bauthor{\bsnm{Sharma}, \binits{V.}}:
\batitle{Smart education with artificial intelligence based determination of
  learning styles}.
\bjtitle{Procedia Computer Science}
\bvolume{132},
\bfpage{834}--\blpage{842}
(\byear{2018})
\doiurl{10.1016/j.procs.2018.05.095}
\end{barticle}
\endbibitem

\bibitem[\protect\citeauthoryear{Corbett and
  Anderson}{1994}]{corbett1994knowledge}
\begin{barticle}
\bauthor{\bsnm{Corbett}, \binits{A.T.}},
\bauthor{\bsnm{Anderson}, \binits{J.R.}}:
\batitle{Knowledge tracing: Modeling the acquisition of procedural knowledge}.
\bjtitle{User Modeling and User-Adapted Interaction}
\bvolume{4}(\bissue{4}),
\bfpage{253}--\blpage{278}
(\byear{1994})
\end{barticle}
\endbibitem

\bibitem[\protect\citeauthoryear{Choi and et~al.}{2020}]{choi2020towards}
\begin{bchapter}
\bauthor{\bsnm{Choi}, \binits{Y.}},
\bauthor{\bsnm{al.}}:
\bctitle{Towards an appropriate query, key, and value computation for knowledge
  tracing}.
In: \bbtitle{Proceedings of the Seventh ACM Conference on Learning @ Scale}
(\byear{2020}).
\doiurl{10.1145/3386527.3405945}
\end{bchapter}
\endbibitem

\bibitem[\protect\citeauthoryear{Essa et~al.}{2023}]{essa2023personalized}
\begin{barticle}
\bauthor{\bsnm{Essa}, \binits{S.G.}},
\bauthor{\bsnm{Celik}, \binits{T.}},
\bauthor{\bsnm{Human-Hendricks}, \binits{N.E.}}:
\batitle{Personalized adaptive learning technologies based on machine learning
  techniques to identify learning styles: A systematic literature review}.
\bjtitle{IEEE Access}
\bvolume{11},
\bfpage{48392}--\blpage{48409}
(\byear{2023})
\doiurl{10.1109/access.2023.3276439}
\end{barticle}
\endbibitem

\bibitem[\protect\citeauthoryear{Gervet et~al.}{2020}]{gervet2020when}
\begin{barticle}
\bauthor{\bsnm{Gervet}, \binits{T.}},
\bauthor{\bsnm{Koedinger}, \binits{K.}},
\bauthor{\bsnm{Schneider}, \binits{J.}},
\bauthor{\bsnm{Mitchell}, \binits{T.}}:
\batitle{When is deep learning the best approach to knowledge tracing?}
\bjtitle{Journal of Educational Data Mining}
\bvolume{12}(\bissue{3}),
\bfpage{31}--\blpage{54}
(\byear{2020})
\doiurl{10.5281/zenodo.4143614}
\end{barticle}
\endbibitem

\bibitem[\protect\citeauthoryear{Liu et~al.}{2023}]{liu2023survey}
\begin{botherref}
\oauthor{\bsnm{Liu}, \binits{Q.}},
\oauthor{\bsnm{Shen}, \binits{S.}},
\oauthor{\bsnm{Huang}, \binits{Z.}},
\oauthor{\bsnm{Chen}, \binits{E.}},
\oauthor{\bsnm{Zheng}, \binits{Y.}}:
A Survey of Knowledge Tracing.
Available: \url{https://arxiv.org/abs/2105.15106}
(2023)
\end{botherref}
\endbibitem

\bibitem[\protect\citeauthoryear{Lee and Yeung}{2019}]{lee2019knowledge}
\begin{botherref}
\oauthor{\bsnm{Lee}, \binits{J.}},
\oauthor{\bsnm{Yeung}, \binits{D.-Y.}}:
Knowledge Query Network for Knowledge Tracing.
arXiv (Cornell University)
(2019).
\doiurl{10.1145/3303772.3303786}
\end{botherref}
\endbibitem

\bibitem[\protect\citeauthoryear{Mao et~al.}{2018}]{mao2018deep}
\begin{barticle}
\bauthor{\bsnm{Mao}, \binits{Y.}},
\bauthor{\bsnm{Lin}, \binits{C.}},
\bauthor{\bsnm{Chi}, \binits{M.}}:
\batitle{Deep learning vs. bayesian knowledge tracing: Student models for
  interventions}.
\bjtitle{Journal of Educational Data Mining}
\bvolume{10}(\bissue{2}),
\bfpage{28}--\blpage{54}
(\byear{2018})
\doiurl{10.5281/zenodo.3554691}
\end{barticle}
\endbibitem

\bibitem[\protect\citeauthoryear{Nakagawa et~al.}{2021}]{nakagawa2019graph}
\begin{barticle}
\bauthor{\bsnm{Nakagawa}, \binits{H.}},
\bauthor{\bsnm{Iwasawa}, \binits{Y.}},
\bauthor{\bsnm{Matsuo}, \binits{Y.}}:
\batitle{Graph-based knowledge tracing: Modeling student proficiency using
  graph neural networks}.
\bjtitle{Web Intelligence}
\bvolume{19}(\bissue{1--2}),
\bfpage{87}--\blpage{102}
(\byear{2021})
\doiurl{10.3233/web-210458}
\end{barticle}
\endbibitem

\bibitem[\protect\citeauthoryear{Piech et~al.}{2015}]{piech2015deep}
\begin{bchapter}
\bauthor{\bsnm{Piech}, \binits{C.}},
\bauthor{\bsnm{Bassen}, \binits{J.}},
\bauthor{\bsnm{Huang}, \binits{J.}},
\bauthor{\bsnm{Ganguli}, \binits{S.}},
\bauthor{\bsnm{Sahami}, \binits{M.}},
\bauthor{\bsnm{Guibas}, \binits{L.J.}},
\bauthor{\bsnm{Sohl-Dickstein}, \binits{J.}}:
\bctitle{Deep knowledge tracing}.
In: \bbtitle{Advances in Neural Information Processing Systems},
pp. \bfpage{505}--\blpage{513}
(\byear{2015})
\end{bchapter}
\endbibitem

\bibitem[\protect\citeauthoryear{Pandey and
  Karypis}{2019}]{pandey2019selfattentive}
\begin{botherref}
\oauthor{\bsnm{Pandey}, \binits{S.}},
\oauthor{\bsnm{Karypis}, \binits{G.}}:
A Self-Attentive Model for Knowledge Tracing.
Available: \url{https://arxiv.org/abs/1907.06837}
(2019)
\end{botherref}
\endbibitem

\bibitem[\protect\citeauthoryear{Pokrajac et~al.}{2016}]{Pokrajac2016}
\begin{bchapter}
\bauthor{\bsnm{Pokrajac}, \binits{D.D.}},
\bauthor{\bsnm{Sudler}, \binits{K.R.}},
\bauthor{\bsnm{Edamatsu}, \binits{P.Y.}},
\bauthor{\bsnm{Hardee}, \binits{T.}}:
\bctitle{Prediction of retention at historically black college/university using
  artificial neural networks}.
In: \bbtitle{2016 IEEE International Conference on Neural Networks (IJCNN)}
(\byear{2016}).
\doiurl{10.1109/neurel.2016.7800124} .
\burl{https://doi.org/10.1109/neurel.2016.7800124}
\end{bchapter}
\endbibitem

\bibitem[\protect\citeauthoryear{Patil and Thorat}{2016}]{patil2016categorical}
\begin{barticle}
\bauthor{\bsnm{Patil}, \binits{S.S.}},
\bauthor{\bsnm{Thorat}, \binits{R.R.}}:
\batitle{Categorical data encoding: Challenges and opportunities}.
\bjtitle{International Journal of Computer Applications}
\bvolume{156}(\bissue{8}),
\bfpage{23}--\blpage{27}
(\byear{2016})
\end{barticle}
\endbibitem

\bibitem[\protect\citeauthoryear{Song et~al.}{2022}]{song2022survey}
\begin{barticle}
\bauthor{\bsnm{Song}, \binits{X.}},
\bauthor{\bsnm{Li}, \binits{J.}},
\bauthor{\bsnm{Cai}, \binits{T.}},
\bauthor{\bsnm{Yang}, \binits{S.}},
\bauthor{\bsnm{Yang}, \binits{T.}},
\bauthor{\bsnm{Liu}, \binits{C.}}:
\batitle{A survey on deep learning based knowledge tracing}.
\bjtitle{Knowledge-Based Systems}
\bvolume{258},
\bfpage{110036}
(\byear{2022})
\doiurl{10.1016/j.knosys.2022.110036}
\end{barticle}
\endbibitem

\bibitem[\protect\citeauthoryear{Wang et~al.}{2023}]{wang2023dynamic}
\begin{botherref}
\oauthor{\bsnm{Wang}, \binits{F.}}, et al.:
Dynamic cognitive diagnosis: An educational priors-enhanced deep knowledge
  tracing perspective.
IEEE Transactions on Learning Technologies,
1--17
(2023)
\doiurl{10.1109/tlt.2023.3254544}
\end{botherref}
\endbibitem

\bibitem[\protect\citeauthoryear{Waheed et~al.}{2020}]{waheed2020predicting}
\begin{barticle}
\bauthor{\bsnm{Waheed}, \binits{H.}},
\bauthor{\bsnm{Hassan}, \binits{S.-U.}},
\bauthor{\bsnm{Aljohani}, \binits{N.R.}},
\bauthor{\bsnm{Hardman}, \binits{J.}},
\bauthor{\bsnm{Alelyani}, \binits{S.}},
\bauthor{\bsnm{Nawaz}, \binits{R.}}:
\batitle{Predicting academic performance of students from vle big data using
  deep learning models}.
\bjtitle{Computers in Human Behavior}
\bvolume{104},
\bfpage{106189}
(\byear{2020})
\doiurl{10.1016/j.chb.2019.106189}
\end{barticle}
\endbibitem

\bibitem[\protect\citeauthoryear{Yağcı}{2022}]{yagci2022educational}
\begin{botherref}
\oauthor{\bsnm{Yağcı}, \binits{M.}}:
Educational data mining: prediction of students’ academic performance using
  machine learning algorithms.
Smart Learning Environments
\textbf{9}(1)
(2022)
\doiurl{10.1186/s40561-022-00192-z}
\end{botherref}
\endbibitem

\bibitem[\protect\citeauthoryear{Yeung and Yeung}{2018}]{yeung2018addressing}
\begin{botherref}
\oauthor{\bsnm{Yeung}, \binits{C.-K.}},
\oauthor{\bsnm{Yeung}, \binits{D.-Y.}}:
Addressing two problems in deep knowledge tracing via prediction-consistent
  regularization.
arXiv
(2018)
{\href{https://arxiv.org/abs/1806.02180}{{arXiv:1806.02180}}}.
Available at \url{https://arxiv.org/abs/1806.02180}
\end{botherref}
\endbibitem

\bibitem[\protect\citeauthoryear{Zine et~al.}{2019}]{zine2019comparative}
\begin{botherref}
\oauthor{\bsnm{Zine}, \binits{O.}},
\oauthor{\bsnm{Derouich}, \binits{A.}},
\oauthor{\bsnm{Talbi}, \binits{A.}}:
A comparative study of the most influential learning styles used in adaptive
  educational environments.
International Journal of Advanced Computer Science and Applications
\textbf{10}(11)
(2019)
\doiurl{10.14569/ijacsa.2019.0101171}
\end{botherref}
\endbibitem

\bibitem[\protect\citeauthoryear{Zhang et~al.}{2017}]{zhang2017dynamic}
\begin{botherref}
\oauthor{\bsnm{Zhang}, \binits{J.}},
\oauthor{\bsnm{Shi}, \binits{X.}},
\oauthor{\bsnm{King}, \binits{I.}},
\oauthor{\bsnm{Yeung}, \binits{D.-Y.}}:
Dynamic key-value memory networks for knowledge tracing.
arXiv
(2017)
{\href{https://arxiv.org/abs/1611.08108}{{arXiv:1611.08108}}}.
Available at \url{https://arxiv.org/abs/1611.08108}
\end{botherref}
\endbibitem

\end{thebibliography}

\end{document}